\documentclass[%
 aip,
 numerical,
 jcp,
rsi,%
amsmath,
amssymb,
preprint,%
author-year,%
]{revtex4-1}
\usepackage{xcolor}
\usepackage{graphicx}
\usepackage{dcolumn}
\usepackage{bm}
\usepackage{natbib}
\usepackage{hyperref}
\usepackage{amsmath}
\usepackage{mathrsfs} 
\usepackage{nicefrac}
\usepackage{mathtools}
\usepackage{bm}

\bibpunct{(}{)}{,}{n}{,}{,}

\begin{document}

\title{Phase-space matrix representation of differential equations for obtaining the energy spectrum of model quantum systems}

\author{Juan C. Morales, Carlos A. Arango}
\email{caarango@icesi.edu.co}
\affiliation{Department of Chemical Sciences, Universidad Icesi, Cali, Colombia}

\date{\today}

\begin{abstract}
Employing the phase-space representation of second order ordinary differential equations we developed a method to find the eigenvalues and eigenfunctions of the 1-dimensional time independent Schrödinger equation for quantum model systems. The method presented simplifies some approaches shown in textbooks, based on asymptotic analyses of the time-independent Schrödinger equation, and power series methods with recurrence relations.  In addition, the method presented here facilitates the understanding of the relationship between the ordinary differential equations of the mathematical physics and the time independent Schrödinger equation of physical models as the harmonic oscillator, the rigid rotor, the Hydrogen atom, and the Morse oscillator.  

\end{abstract}


\maketitle 

\section{Introduction}

The 1-dimensional time independent Schrödinger equation (TISE) can be solved analytically for few physical models. The harmonic oscillator, the rigid rotor, the Hydrogen atom, and the Morse oscillator are examples of physical models with known analytical solution of the TISE \cite{Atkins2011}. The analytical solution of the TISE for a physical model is usually obtained by using the ansatz of a wavefunction as a product of two functions, one of these functions acts as an integrating factor \cite{Derrick1997}, the other function produces a differential equation solvable either by Frobenius series method or by directly comparing with a template ordinary differential equation (ODE) with known solution \cite{Morse1953}. Examples of template ODEs of physical interest are the Hermite, Laguerre, Legendre, and confluent hypergeometric equations. Before using the wavefunction product ansatz is necessary to write the TISE in terms of a non-dimensional coordinate, and make a transformation of coordinates that depends on the potential energy function. It is important to mention that not always straightforward to find a product wavefunction with an integrating factor that produces a TISE easy to solve by Frobenuis method or comparable with a template ODE with known solution.

The analysis in phase-space of second order ODEs uses to geometrical and mathematical tools that allow to understand qualitatively the behaviour of the solutions of the ODE without having an exact solution \cite{Falco1976,Arnold1992,Hubbard1994}. The solutions to a 1-dimensional second order ODE in phase-space are trajectories in the 2-dimensional phase-space instead of the 1-dimensional function obtained by solving the ODE in space representation. A stability analysis of the tangent phase-space vector field of an ODE allows to find and classify fixed points, and bound phase-space trajectories \cite{Strogatz1994}. The phase-space analysis of the TISE gives the bound states of a quantum systems in terms of the fixed points and the bound trajectories \cite{Molano2021}. 

A 1-dimensional second order ODE can be described in phase-space as a couple of first order ODEs. This set of ODEs in phase-space can be written in matrix fashion by defining a phase-space vector as the concatenation of the function and its space derivative. The resulting matrix equation relates the tangent to the phase-space vector with the product of a $2\times 2$ matrix times the phase-space vector \cite{Arnold1992}. Since the TISE for a physical system depends on a variable which is a physical quantity, with dimensions of length, it is necessary to define a dimensionless variable before using integrating factors and transformation of coordinates to obtain an equation comparable with the template ODEs \cite{Langtangen2016}. As a second order ODE, the TISE can be represented in terms of the phase-space wavefunction vector. The matrix representation of the TISE in phase-space allows to use integrating factors that make the matrix form of the TISE comparable with the matrix form of the template ODE. In this work we introduce a method based on the phase-space representation of both, the TISE, and a template second order ODE. The method allows to obtain directly nondimensionalization constants, integrating factors, and the solution of the TISE for model quantum systems. Instead of using a comparison term by term between the TISE and the template ODE, the method of this work is based in an algebraic equation obtained by comparing the phase-space matrix representation of the TISE and a template ODE with known solution in the form of orthonormal polynomials. Section \ref{sec:applications} of this article displays the use of the equations deduced in section \ref{sec:theory} to calculate the energy spectrum and the eigenfunctions of model quantum systems: the harmonic oscillator, the rigid rotor, the hydrogen atom, and the Morse oscillator. 

The method developed and presented in this article is limited to finding analytical solutions of 1-dimensional quantum systems, of academic interest, in terms of orthogonal polynomials. A generalization of this method has been used to obtain numerically bound and resonance states of the rotationally excited Hydrogen molecule\cite{Molano2021}. There are different approaches to find eigenvalues of the TISE using phase-space representation. Phase-space integration based in the Wentzel–Kramers–Brillouin
(WKB) method has been used to calculate accurately the eigenvalues of anharmonic oscillators, and square and logartihmic potentials \cite{Nagabhushana1997,Mei1998}. Primitive semi-classical methods that use phase-space integration based in the Wilson-Sommerfeld quantization rules provide excellent results for anharmonic oscillators \cite{Mukhopadhyay2001}. 

\section{Theory and Methods}\label{sec:theory}

\subsection{Phase-space representation of the time independent Schr\"odinger equation}

A homogeneous second order linear differential equation, $P(x)y''(x)+Q(x)y'(x)+R(x)y(x)=0$, can be written in terms of the phase-space vector $\mathbf{y}(x)=\begin{psmallmatrix} y(x) \\ y'(x)\end{psmallmatrix}$, obtaining a matrix equation $\mathbf{y}'(x)=\mathsf{D}(x)\mathbf{y}(x)$ with matrix $\mathsf{D}=\mathsf{D}(x)$ given by
\begin{equation}\label{eq:D_matrix_def}
    \mathsf{D}=\begin{pmatrix}0 & 1 \\
    -\frac{R(x)}{P(x)} & -\frac{Q(x)}{P(x)}
    \end{pmatrix}.
\end{equation}

The 1-dimensional time independent Schr\"odinger equation (TISE) is given by
\begin{equation}\label{eq:TISE}
    -\frac{\hbar^2}{2m}\psi''(q)+B(q)\psi'(q)+V(q)\psi=E\psi(q),
\end{equation}
The physical coordinate $q$ can be written as $q=x_c x$, with $x$ as a dimensionless coordinate, and $x_c$ a constant carrying out the length dimension. The dimensionless TISE is given by
\begin{equation}
    -\varphi''(x)+b(x)\varphi'(x)+v(x)\varphi(x)=\varepsilon\varphi(x), 
\end{equation}
with $\varphi(x)=\psi(q)$, $b(x)=\frac{B(q)}{a x_c}$, $v(x)=\frac{V(q)}{a}$, $\varepsilon=\frac{E}{a}$, and $a=\frac{\hbar^2}{2 m x_c^2}$. The dimensionless TISE can be written in matrix fashion by using the phase-space wavefunction vector $\bm{\varphi}=\begin{psmallmatrix}\varphi \\ \varphi' \end{psmallmatrix}$, 
\begin{equation}\label{eq:phase_space_TISE}
    \bm{\varphi}'=\mathsf{A}\bm{\varphi},
\end{equation}
with $\mathsf{A}=\mathsf{A}(x)$ is given by
\begin{equation}\label{eq:A_matrix_def}
    \mathsf{A}=\begin{pmatrix}
    0 & 1 \\
    -k^2(x) & b(x)
    \end{pmatrix}, 
\end{equation}
and $k^2(x)=\left(\varepsilon-v(x)\right)$.

It is convenient to write the wave function $\varphi$ as a product $\varphi=fg$ of two functions $f=f(x)$ and $g=g(x)$, with derivative $\varphi'=\left(f'g+fg'\right)$. The product function $\varphi=fg$, and its derivative give a matrix equation for $\bm{\varphi}$ in terms of the vector  $\mathbf{f}=\begin{psmallmatrix} f \\ f' \end{psmallmatrix}$,
\begin{equation}\label{eq:Psi_of_F}
    \bm{\varphi}=\mathsf{B}\mathbf{f},
\end{equation}
with matrix $\mathsf{B}=\mathsf{B}(x)$ given by
\begin{equation}\label{eq:B_matrix_def}
    \mathsf{B}=\begin{pmatrix}
    g & 0 \\
    g' & g
    \end{pmatrix}. 
\end{equation}
The $x$-derivative of $\bm{\varphi}=\mathsf{B}\mathbf{f}$ is given by
\begin{equation}\label{eq:Psi_prime_of_F}
    \bm{\varphi}'=\mathsf{B}'\mathbf{f}+\mathsf{B}\mathbf{f}',
\end{equation}
with
\begin{equation}\label{eq:B_matrix_deriv}
    \mathsf{B}'=\begin{pmatrix}
    g' & 0 \\
    g'' & g' 
    \end{pmatrix}.
\end{equation}
The use of $\bm{\varphi}=\mathsf{B}\mathbf{f}$ in equation \eqref{eq:phase_space_TISE} gives $\bm{\varphi}'=\mathsf{AB}\mathbf{f}$. This result for $\bm{\varphi}'$ can be used in the left-hand-side of equation \eqref{eq:Psi_prime_of_F} to obtain $\mathsf{AB}\mathbf{f}=\mathsf{B}'\mathbf{f}+\mathsf{B}\mathbf{f}'$, which can be solved for $\mathbf{f}'$ giving 
\begin{equation}\label{eq:F_deriv}
\begin{split}
    \mathbf{f}'&=\mathsf{B}^{-1}\left(\mathsf{AB}-\mathsf{B}'\right)\mathbf{f}\\
    &=\mathsf{C}\mathbf{f},
\end{split}
\end{equation}
with the matrix $\mathsf{C}=\mathsf{B}^{-1}\left(\mathsf{AB}-\mathsf{B}'\right)$ given explicitly by
\begin{equation}\label{eq:C_matrix_def}
    \mathsf{C}=\begin{pmatrix}
    0 & 1 \\
    -\frac{gk^2-bg'+g''}{g} & b-\frac{2g'}{g}
    \end{pmatrix}.
\end{equation}

Matrices $\mathsf{C}$ and $\mathsf{D}$, equation \eqref{eq:C_matrix_def} and \eqref{eq:D_matrix_def} respectively, can be equaled to obtain a set of ODEs for the integrating factor $g$
\begin{align}
    -\frac{gk^2-bg'+g''}{g}&=-\frac{R}{P},\label{eq:ODEs_1}\\
    b-\frac{2g'}{g}&=-\frac{Q}{P}.\label{eq:ODEs_2}
\end{align}
Integration of equation \eqref{eq:ODEs_2} gives the function $g$ in terms of the known functions $P$, $Q$ and $b$,
\begin{equation}\label{eq:g_function}
    g=\exp{\left(\int{\frac{Q+bP}{2P}dx}\right)}.
\end{equation}

Equations \eqref{eq:ODEs_1} and \eqref{eq:ODEs_2} are combined to obtain the algebraic equation
\begin{equation}\label{eq:PQR_eqn}
    k^2+\frac{b'}{2}-\frac{b^2}{4}=G(x),
\end{equation}
with $G(x)=-\frac{Q^2-2QP'+2P(Q'-2R)}{4P^2}$. Since $k^2$ is a function of the energy $E$, equation \eqref{eq:PQR_eqn} can be used to obtain the energy eigenvalues of the TISE. The left hand side (LHS) of equation \eqref{eq:PQR_eqn} depends on the coefficients of the TISE, meanwhile the right hand side (RHS) depends on the coefficients of a template equation. Examples of template ODEs employed in physics with the respective $G$ functions are shown in the second line of the first column and the second column, respectively, of table \ref{table:template_ODE_and_G}.

\begin{table}
\begin{tabular}{ ll } 
Template ODE & $G$ function \\
\hline \hline
Hermite &       \\
$y''-2xy'+2\lambda y=0$ \hspace{1cm} & $G(x)=1+2\lambda-x^2$ \\
$x\in(-\infty,\infty)$,\hspace{0.5cm}$y=H_\lambda(x)$; \hspace{0.5cm}$\lambda\in\mathbb{Z}$, $\lambda\ge 0$ & \\
\hline
Associated Legendre   &  \\ 
$(1-x^2)y''-2xy'+\left(l(l+1)-\frac{m^2}{1-x^2}\right)y=0$ \hspace{1cm} & $G(x)=-\frac{m^2-1+(x^2-1)(l+1)l}{(x^2-1)^2}$\\
$x\in[-1,1]$,\hspace{0.5cm}$y=P_l^m(x)$; \hspace{0.5cm}$l,m\in\mathbb{Z}$, $-l\le m \le l$ & \\
\hline 
Polar Associated Legendre & \\   
$ y''(\theta)+\cot{\theta}y'(\theta)+\left(l(l+1)-\frac{m^2}{\sin^2{\theta}}\right)y(\theta)=0 $ \hspace{1cm} & $G(\theta)=\frac{1}{4}+l(l+1)+\frac{\left(\nicefrac{1}{4}-m^2\right)}{\sin^{2}{\theta}} $ \\
$y=P_l^m(\cos{\theta})$; \hspace{0.5cm} $l,m\in\mathbb{Z}$, $-l\le m \le l$ & \\
\hline Associated Laguerre  &   \\ 
$xy''+(\nu+1-x)y'+\lambda y=0$ \hspace{1cm}&  $G(x)=-\frac{1}{4}+\frac{1+\nu+2\lambda}{2x}+\frac{1-\nu^2}{4x^2}$ \\
$x\in[0,\infty)$,\hspace{0.5cm}$y=L_\lambda^\nu(x)$; \hspace{0.5cm} $\lambda,\nu\in\mathbb{Z}$, $\lambda\geq 0$, $\nu>0$ \\
\hline 
Confluent Hypergeometric   &  \\
$xy''+(c-x)y'-ay=0$  \hspace{1cm} & $G(x)=-\frac{1}{4}+\frac{c-2a}{2x}+\frac{c(2-c)}{4x^2}$ \\
$\mathstrut_1 F_1(a,c,x)$; \hspace{0.5cm} $a\in\mathbb{Z}$, $a\le 0$, $c>0$ \\
\hline\hline
\end{tabular}
\caption{Examples of template ODEs used in physics with their respective polynomial solutions, and $G$ functions.}
\label{table:template_ODE_and_G}
\end{table}

The template ODEs $P(x)y''(x)+Q(x)y'(x)+R(x)y(x)=0$ usually  have polynomial solutions $y(x)=p_{\lambda,\nu}(x)$ of degree $\lambda$ and order $\nu$. Examples polynomial solutions for template ODEs, and the mathematical conditions for the polynomial solutions, are given in the third line of the first column of table \ref{table:template_ODE_and_G}. The integrating factor $g(x)$ of equation \eqref{eq:g_function}, and the polynomial solutions $p_{\lambda,\nu}(x)$ of the template ODE give the solution of the TISE
\begin{equation}
    \psi(q(x))=N_{\lambda,\nu}g(x)p_{\lambda,\nu}(x),
\end{equation}
with $N_{\lambda,\nu}$ as a normalization constant.

\section{Applications}\label{sec:applications}

\subsection{The harmonic oscillator}

In terms of the dimensionless coordinate $x\in\mathbb{R}$, the TISE for the harmonic oscillator
\begin{equation}
-\frac{\hbar^2}{2m x_c^2}\varphi''(x)+\tfrac{1}{2}m\omega^2x_c^2x^2\varphi(x)=E\varphi(x),
\end{equation}
has $b=0$, and
\begin{equation}\label{eq:HO_k}
    k^2=\frac{2m x_c^2}{\hbar^2}\left(E-\tfrac{1}{2}m\omega^2x_c^2x^2\right).
\end{equation}
The value of $k^2$ of equation \eqref{eq:HO_k} can be used on the LHS of equation \eqref{eq:PQR_eqn} producing a quadratic term in $x$ that can be cancelled only by using the Hermite $G(x)$ of table \ref{table:template_ODE_and_G} ,
\begin{equation}\label{eq:harmonic_oscillator_k2}
        \frac{2m x_c^2}{\hbar^2}\left(E-\tfrac{1}{2}m\omega^2x_c^2x^2\right)=1+2\lambda-x^2.
\end{equation}
The equality in \eqref{eq:harmonic_oscillator_k2} is hold if
\begin{align}
    \frac{2m x_c^2}{\hbar^2}E&=1+2\lambda, \label{eq:HO_k1}\\
    -\frac{m^2 \omega^2 x_c^4}{\hbar^2}x^2&=-x^2. \label{eq:HO_k2}
\end{align}
Equation \eqref{eq:HO_k2} gives the constant $x_c=\left(\frac{\hbar}{m\omega}\right)^{\nicefrac{1}{2}}$. Substitution of this value of $x_c$ in equation \eqref{eq:HO_k1} gives the energy $E=\hbar\omega\left(\lambda+\nicefrac{1}{2}\right)$. 

The integrating factor $g$ of equation \eqref{eq:g_function} can be obtained for the harmonic oscillator and the Hermite equation, $g(x)=e^{-x^2/2}$.  Bound solutions of the Hermite differential equation are given by the Hermite polynomials $y=H_\lambda(x)$ with $\lambda\in\mathbb{Z}$ and $\lambda\geq 0$. The bound solution of the Harmonic oscillator is given by
\begin{equation}
    \psi_\lambda(q)=N_\lambda\exp{\left(-\tfrac{q^2}{2 x_c^2}\right)} H_\lambda\left(\tfrac{q}{x_c}\right),
\end{equation}
with $N_\lambda$ as a normalization constant.

\subsection{The Rigid Rotor}

In spherical coordinates, the Schr\"odinger equation for a rigid rotor is given by
\begin{equation}\label{eq:TISE_rigid_rotor}
-\frac{\hbar^2}{2 \mu R^2}\left(\frac{1}{\sin{\theta}}\left(\sin{\theta}\frac{\partial}{\partial\theta}\right)+\frac{1}{\sin^2{\theta}}\frac{\partial^2}{\partial\phi^2}\right)\psi(\theta,\phi)=E\psi(\theta,\phi),
\end{equation}
with $\mu$ and $R$ as the reduced mass and bond length of the rotor, respectively. The use of a wave function as a product $\psi(\theta,\phi)=\Theta(\theta)\Phi(\phi)$ gives
\begin{align}\label{eq:TISE_Theta}
    \frac{1}{\Theta}\sin{\theta}\frac{d}{d\theta}\left(\sin{\theta\frac{d\Theta}{d\theta}}\right)+\beta\sin^2{\theta}&=C,\\
    -\frac{1}{\Phi}\frac{d^2\Phi}{d\phi^2}&=C,\label{eq:TISE_Phi}
\end{align}
with $C$ as a constant and the dimensionless $\beta={2\mu R^2 E}/{\hbar^2}$. Normalized $2\pi$-periodic solutions of \eqref{eq:TISE_Phi} are given by $\Phi(\phi)=\frac{1}{\sqrt{2\pi}}e^{i m\phi}$ with $m\in\mathbb{Z}$. The use of $\Phi(\phi)=\frac{1}{\sqrt{2\pi}}e^{i m\phi}$ in equation \eqref{eq:TISE_Phi} produces $C=m^2$; the use of this $C$ in equation \eqref{eq:TISE_Theta} produces
\begin{equation}\label{eq:TISE_Theta_2}
 \frac{d^2\Theta}{d\theta^2}+\cot{\theta}\frac{d\Theta}{d\theta}+\frac{\beta \sin^2{\theta}-m^2}{\sin^2{\theta}}\Theta=0.
 \end{equation}

The use of the vector $\mathbf{\Theta}=\{\Theta,\Theta'\}$ allows to write this equation in matrix fashion, $\mathbf{\Theta}'=\mathsf{A}(\theta)\mathbf{\Theta}$. The matrix $\mathsf{A}$ has exactly the same form of equation \eqref{eq:A_matrix_def} with $b=-\cot{\theta}$ and $k^2=\frac{\beta \sin^2{\theta}-m^2}{\sin^2{\theta}}$. The use of these values of $b$ and $k^2$ on the RHS of equation \eqref{eq:PQR_eqn} gives
\begin{equation}
\frac{1}{4}+\beta+\frac{\nicefrac{1}{4}-m^2}{\sin^2{\theta}}=G(\theta).
\end{equation}
The LHS of this equation matches with the $G(\theta)$ of the polar form of the Associated Legendre equation with $\beta=l(l+1)$. Recalling that $\beta=\frac{2IE}{\hbar^2}$, the energy of the rigid rotor is given by $E_l=\frac{\hbar^2}{2I}l(l+1)$. Finally, the use of equation \eqref{eq:g_function} gives 
\begin{equation}
    g(\theta)=1.
\end{equation}
The solution of equation \eqref{eq:TISE_Theta_2} is given by $\Theta_{l,m}(\theta)=N_{l,m}P_{l}^{m}(\cos{\theta})$ with $P_{l}^{m}$ as an associated Legendre polynomial, and $N_{l,m}$ a normalization constant. The full solution for the rigid rotor, equation \eqref{eq:TISE_rigid_rotor}, is given by 
\begin{equation}
\begin{split}
    \psi(\theta,\phi)&=N_{l,m}P_{l}^{m}(\cos{\theta})e^{i m\phi}\\
    &=Y_{l}^{m}(\theta,\phi),
\end{split}
\end{equation}
with $Y_{l}^{m}(\theta,\phi)$ as the spherical harmonic functions.

\subsection{The radial equation for the Hydrogen atom}

In terms of the dimensionless coordinate $\rho$, related to the physical coordinate $r$ by $r=r_c\rho$, the dimensionless radial Schr\"odinger equation for Hydrogen is given by
\begin{equation}\label{eq:TISE_H_atom}
    R''(\rho)+\frac{2}{\rho}R'(\rho)+k_l^2(\rho)R(\rho)=0,
\end{equation}
having
\begin{equation}\label{eq:k2_H_atom}
    k_l^2=-\frac{r_c^2 E}{a_0^2 E_g}+\frac{2r_c}{a_0 \rho}-\frac{l(l+1)}{\rho^2},
\end{equation}
with $l$ a non-negative integer, $a_0=\frac{4\pi\epsilon_0\hbar^2}{m_e e^2}$, and $E_g=-\frac{\hbar^2}{2m_e a_0^2}$, as the Bohr radius and ground state energy, respectively. The coefficient of the first derivative of equation \eqref{eq:TISE_H_atom} gives $b=-\nicefrac{2}{\rho}$.  The use of $b=-\nicefrac{2}{\rho}$ and the Associated Laguerre $G(\rho)$ of table \ref{table:template_ODE_and_G} in equation  \eqref{eq:PQR_eqn} gives
\begin{equation}\label{eq:k2_vs_G_H_atom}
    -\frac{r_c^2 E}{a_0^2 E_g}+\frac{2r_c}{a_0 \rho}-\frac{l(l+1)}{\rho^2}=-\frac{1}{4}+\frac{1-\nu^2}{4\rho^2}+\frac{1+\nu+2\lambda}{2\rho}.
\end{equation}
The equality of \eqref{eq:k2_vs_G_H_atom} leads to
\begin{align}
    -\frac{r_c^2 E}{a_0^2 E_g}&=-\frac{1}{4}, \label{eq:k2_vs_G_Hatom_eqn1}\\
    \frac{2r_c}{a_0 \rho}&=\frac{1+\nu+2\lambda}{2\rho}, \label{eq:k2_vs_G_H_atom_eqn2} \\
    -\frac{l(l+1)}{\rho^2}&=\frac{1-\nu^2}{4\rho^2}.\label{eq:k2_vs_G_H_atom_eqn3}
\end{align}
Equation \eqref{eq:k2_vs_G_Hatom_eqn1} gives $r_c=\left(\frac{a_0^2 E_g}{4 E}\right)^{\nicefrac{1}{2}}$, equation \eqref{eq:k2_vs_G_H_atom_eqn3} gives $\nu=2l+1$. Finally, using the results of $r_c$ and $\nu$ in equation \eqref{eq:k2_vs_G_H_atom_eqn2} produces $E=\frac{E_g}{n^2}$ with $n=1+l+\lambda$. Using the quantized value of  the energy, $E=\frac{E_g}{n^2}$, the constant $r_c$ changes to $r_c=\frac{a_0 n}{2}$. 

Polynomial solutions of the associated Laguerre ODE are obtained for $\lambda$ and $\nu$ non negative integers, since $l$ is a non-negative integer, $n=1+l+\lambda$ must be a positive integer, and $l$ can take any value $0\le l\le n-1$.

Finally, by using equation \eqref{eq:g_function} the integrating factor $g$ results in $g(\rho)=\rho^{l}e^{-\rho/2}$, which gives for the Hydrogen atom wave functions
\begin{equation}
    R_{n,l}(\tfrac{2r}{n a_0})=N_{n,l}\left(\frac{2r}{n a_0}\right)^{l}e^{-\tfrac{r}{n a_0}}L_{n-l-1}^{2l+1}(\tfrac{2r}{n a_0}),
\end{equation}
with $L_{n-l-1}^{2l+1}$ as an associated Laguerre polynomial, and $N_{n,l}$ as a normalization constant.


\subsection{The Morse Oscillator}

The dimensionless TISE for the Morse potential is given by
\begin{equation}\label{eq:Morse_TISE}
   -\tfrac{\hbar^2}{2m x_c^2}\varphi''(x)+D_e\left(e^{-2\alpha x_c x}-2e^{-\alpha x_c x}\right)\varphi(x)=E\varphi(x),
\end{equation}
with $x \in \mathbb{R}$, $\alpha>0$, and $D_e>0$. The transformation of coordinates $y=\frac{2\sqrt{2mD_e}}{\alpha\hbar}e^{-\alpha x_c x}$ gives 
\begin{equation}\label{eq:Morse_dimensionless_TISE}
\chi''(y)+\frac{1}{y}\chi'(y)+\left(\frac{\varepsilon}{y^2}+\frac{\delta}{y}-\frac{1}{4}\right)\chi(y)=0,
\end{equation}
with $y>0$, $\chi(y)=\varphi(x)$, $\varepsilon=\frac{2mE}{\alpha^2\hbar^2}$, and $\delta=\frac{\sqrt{2mD_e}}{\alpha\hbar}$. The dimensionless TISE \eqref{eq:Morse_dimensionless_TISE} leads to $k^2=\left(\frac{\varepsilon}{y^2}+\frac{\delta}{y}-\frac{1}{4}\right)$, and $b(y)=-1/y$. The use of this values of $b$ and $k^2$ in the LHS of equation \eqref{eq:PQR_eqn}, and the confluent hypergeometric $G$, in the RHS of equation \eqref{eq:PQR_eqn} produces
\begin{equation}\label{eq:PQR_eqn_Morse}
    \frac{1+4\varepsilon}{4y^2}+\frac{\delta}{y}-\frac{1}{4}=-\frac{1}{4}+\frac{c-2a}{2y}+\frac{c(2-c)}{4y^2}.
\end{equation}
The equality \eqref{eq:PQR_eqn_Morse} holds if
\begin{align}
    1+4\varepsilon&=c(2-c),\label{eq:PQR_eqn_Morse_1}\\
    \delta&=\frac{c-2a}{2}.\label{eq:PQR_eqn_Morse_2}
\end{align}
Equations \eqref{eq:PQR_eqn_Morse_1} and \eqref{eq:PQR_eqn_Morse_2} give $a=\tfrac{1}{2}-\delta\pm\sqrt{-\varepsilon}$ and $c=1\pm2\sqrt{-\varepsilon}$. Polynomial solutions of the confluent hypergeometric differential equation are obtained for non positive integer $a$ and nonegative $c$. The number of bound solutions must increase as the value of $\delta$ increases. These requirements are fulfilled by $a=\tfrac{1}{2}-\delta+\sqrt{-\varepsilon}$ and $c=1+2\sqrt{-\varepsilon}$, which give $\varepsilon=-\left(\delta+\left(a-\tfrac{1}{2}\right)\right)^2$, and the energy eigenvalues are
\begin{equation}
    E_n=-D_e\left(1-\frac{n+\nicefrac{1}{2}}{\delta}\right)^2,
\end{equation}
with $n=-a$. Bound states of the Morse oscillator have nonegative integer values of $n$ such that $E_n>-D_e$.

The integrating factor $g=g(y)$ of equation \eqref{eq:g_function} is given by $g=e^{-y/2}y^{\sqrt{-\varepsilon}}$, and the full Morse wavefunction is
\begin{equation}
    \psi(y)=N_{n,c}e^{-y/2}y^{\sqrt{-\varepsilon}}\mathstrut_1 F_1(n,c,y),
\end{equation}
with $y=\frac{2\sqrt{2mD_e}}{\alpha\hbar}e^{-\alpha x_c x}$, $c=1+2\sqrt{-\varepsilon}$, $N_{n,c}$ a normalization constant, and $\mathstrut_1 F_1(n,c,y)$ the confluent hypergeometric function of the first kind.

\section{Conclusions}
The results obtained in section \ref{sec:applications} have shown that the equations developed in section \ref{sec:theory} work exactly for model quantum systems. Equations \eqref{eq:g_function} and \eqref{eq:PQR_eqn} establish a connection between the TISE of a model system and a template ODE, these equations only require the coefficient functions of the TISE and the template ODE. The method presented here avoid the finding of nondimensionalization parameters and integrating factors based on asymptotic analyses of the TISE. The integrating factors are obtained directly from the equations of this method by using a simple factorization ansatz of the wavefunction. In general terms, the use of the method presented in this work facilitates the understanding of the relationship between the ODEs of the mathematical physics and the TISE for model quantum systems. The method presented in this work avoids the use of Frobenius or algebraic ladder operator methods to obtain the energy spectrum of the TISE, instead, it resorts to elements from algebra, linear algebra, and integral and differential calculus, that are easier to use and reach by undergraduate students in courses of Quantum Mechanics or Physical Chemistry.

\section{Acknowledgments}
This project has been fully financed by the internal research grants of University Icesi.
\bibliographystyle{aipnum4-1}
\bibliography{references}
\end{document}